# Correlation Lengths of Red and Blue Galaxies: A New Cosmic Ruler

Michael J. Longo
University of Michigan, Ann Arbor, MI 48109

A comparison of the correlation lengths of red galaxies with blue can provide a new cosmic ruler. Using 269,000 galaxies from the SDSS DR6 survey, I show that the 3D correlation length averaged over many clusters remains very nearly constant at $L_0=4.797\pm0.024$ Mpc/h from small redshifts out to redshifts of 0.5. This serves as a new measure of cosmic length scales as well as a means of testing the standard cosmological model that is almost free of selection biases. The cluster number density also appears to remain constant over this redshift range.

## 1. INTRODUCTION

One of the big challenges in modern cosmology is to calibrate the cosmic distance scale. This has to be done stepwise starting from stars within our Galaxy, then on to nearby galaxies, and eventually out to the farthest structures that are observable. For redshifts $z \leq 0.01$ or so the main standard for the extragalactic distance scale comes from Cepheid variables[1]. These provide an accuracy ranging from 7% for nearby ones to 15% for distant ones. Type Ia supernovae can be used as a distance scale out to $z \sim 5$ with an error approaching 5%.[2] All of the techniques for establishing extragalactic distances involve significant assumptions and generally suffer from poor statistics.

"Old" red (mostly elliptical) galaxies are known to cluster more strongly than younger blue (spiral and irregular) galaxies.[3] This clustering reflects the density fluctuations in the early universe. Presumably the length scales associated with this clustering are constant everywhere and evolve slowly with time or redshift. Thus the average size or correlation length of the clusters can be used as a cosmic ruler that can be "calibrated" against nearby galaxies. This also provides another way to determine the Hubble constant and its possible variation with position. As discussed below, the 3D correlation length can be measured to a precision $\approx 2\%$ out to $z \sim 0.5$. Cluster sizes and counts are an important tool for understanding dark energy and other questions in cosmology.[4] Techniques that require explicit cluster finding are subject to serious biases.[5] The technique described here does not require cluster finding and appears to be almost bias free. The only assumption is that when averaged over sufficiently large volumes of space the clustering length of galaxies is constant or evolves slowly with $z$ for $z < 1$.

The SDSS DR6 database[6] contains ~680,000 galaxies with spectra for $z<1.0$. The sky coverage of the spectroscopic data is almost complete for right ascensions ($RA$) between 120° and 250° and declinations ($\delta$) between -10° and 65°. For the hemisphere toward $RA=0°$, only narrow bands in $\delta$ near -10°, 0°, and +10° are covered. The following discussion will be limited to the hemisphere toward $RA=180°$ except as noted.

## 2. THE SAMPLE

All objects classified as "galaxies" in the SDSS DR6 database were used in this analysis. The definition of "red" and "blue" galaxies used here is based on the difference in luminosities in the ultraviolet ($U$) and far infrared ($Z$) optical bands.[7] Figure 1 shows examples of the ($U$-$Z$) distribution for 3 of the redshift ranges. For small $z$ (Fig. 1a), the distribution shows two fairly well separated peaks corresponding to the bluer spiral and irregular galaxies on the left and the redder ellipticals on the right. For $z\sim0.09$ (Fig. 1b), the distribution has shifted somewhat toward the right due to overall reddening of the spectra, and the spiral peak is much diminished compared to the ellipticals. For larger $z$ (Fig. 1c), the distribution is dominated by a large peak near ($U$-$Z$)=0.5. This is due to the misidentification of quasars as "galaxies" in the sample. For large $z$, it is difficult to distinguish one from the other. The contamination of the galaxy sample with quasars was negligible for $z<0.3$. For $z>0.3$ the quasar positions were found to be correlated like blue galaxies, and they were therefore considered as part of the blue galaxy sample in determining correlation lengths.

To determine the correlation lengths of the elliptical (red) galaxies we can compare their correlation functions to those of the blue galaxies that do not exhibit strong clustering. Here the definition of red and blue is somewhat arbitrary. It can be chosen empirically to maximize the correlations at small separations. For $z< 0.15$ the median $U$-$Z$ was used as the division between red and blue. Galaxies below the median were considered to be "blue" and those above "red". For higher redshifts for which the distinction between spirals and ellipticals is not apparent from the $U$-$Z$ plot, the division point was chosen to maximize the correlation amplitude at small separations. The divisions for the 3 examples in Fig. 1 are indicated.

The correlation lengths derived in what follows were found to be almost independent of the choice of the red-blue division. Errors due to the choice are discussed in the Uncertainties section. In the following discussion I simply refer to the samples as "red" and "blue".

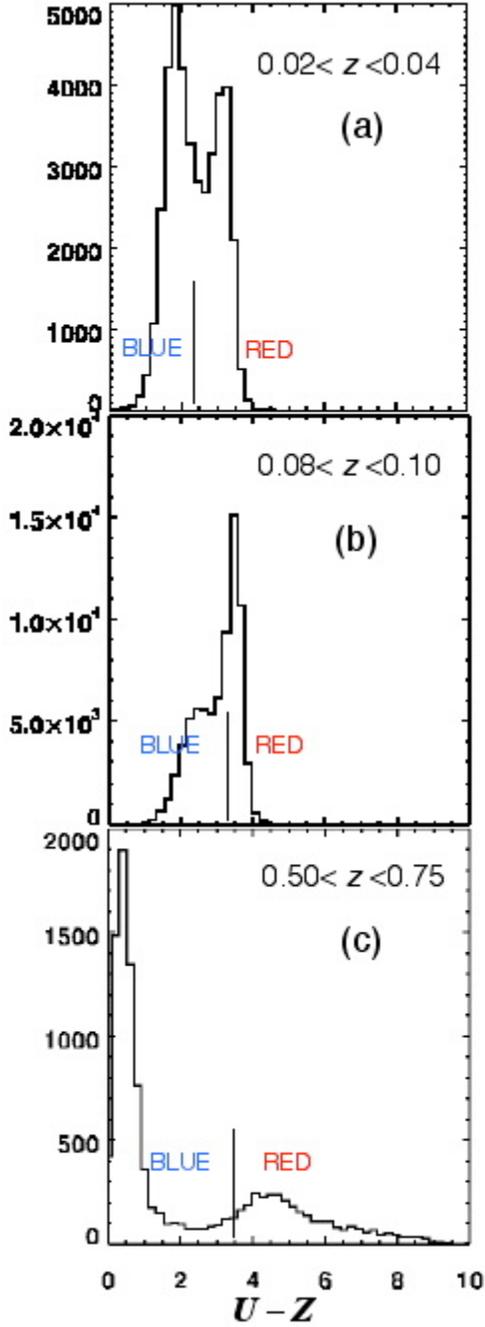 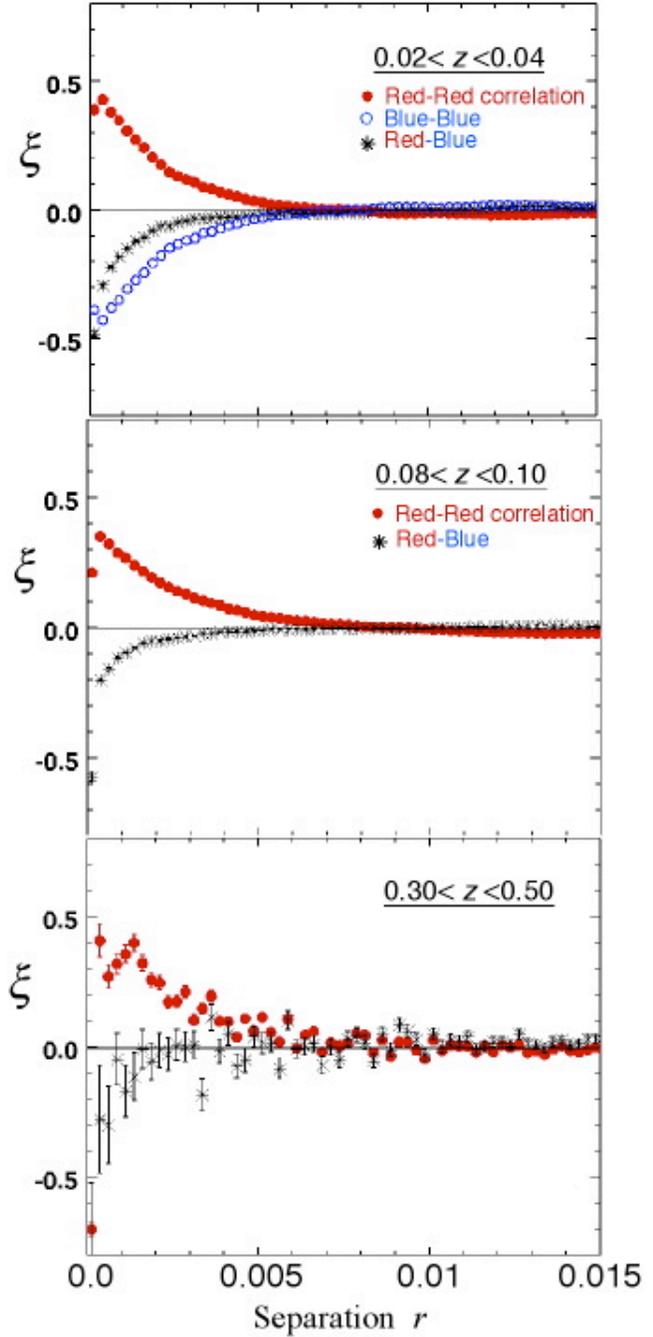

FIG. 1  *U-Z* distributions for 3 redshift ranges. The vertical lines show the division between "red" and "blue" galaxies.

FIG. 2  Correlation *vs.* separation for 3 redshift ranges. The red galaxies show a strong correlation at small separations and a strong anticorrelation with the blue. The error bars are usually smaller than the points.

## 3. CORRELATION FUNCTIONS

Three-dimensional 2-point correlation functions for the galaxies in the red and blue samples were used to determine the cluster correlation lengths. For every pair of galaxies the dimensionless comoving coordinates of each galaxy were calculated. Assuming a Friedmann universe with deceleration parameter $q_0=0$, the comoving coordinates $s_1$ and $s_2$ are given by[8]

$$s_1 = z_1(1+\tfrac{1}{2}z_1)/(1+z_1), \qquad s_2 = z_2(1+\tfrac{1}{2}z_2)/(1+z_2) \qquad (1)$$

The comoving distance between the two galaxies is then the magnitude of the vector

$$\vec{r}' = (s_2 \sin\theta, 0, s_2 \cos\theta) + (0, 0, -s_1)\left\{(1+s_2^2)^{\frac{1}{2}} + \frac{s_2}{s_1}\cos\theta - \frac{s_2}{s_1}\cos\theta(1+s_1^2)^{\frac{1}{2}}\right\} \qquad (2)$$

in units of $c/H_0$ where $H_0$ is Hubble's constant and $\theta$ is the angular separation of the two galaxies. In the following the separations $r$ will be calculated from Eq. 2. As discussed below, the correlation lengths are very insensitive to the choice of $q_0$.

I define the correlation functions used here as

$$\xi_{RR}(r) = N_{RR}/\bar{N} - 1; \qquad \xi_{BB}(r) = N_{BB}/\bar{N} - 1; \qquad \xi_{RB}(r) = N_{RB}/\bar{N} - 1 \qquad (3)$$

where $N_{RR}$ is the number of red-red galaxy pairs in a particular separation bin at $r$; $N_{BB}$ is the number of blue-blue pairs; $N_{RB}$ is the number of red-blue pairs; and $\bar{N}$ is the mean of $N_{RR}$ and $N_{BB}$. Each of the $N$'s was normalized to the total number of pairs summed over all separation bins as defined below. The uncertainties in the $\xi$ are calculated as[9]

$$\Delta\xi = \sqrt{(1+\xi(r))/N} \cong 1/\sqrt{N} \qquad (4)$$

with $N$ being the unnormalized number of pairs defined above with the appropriate subscripts.

The correlations were studied in redshift slices between 0.01 and 0.75. For each $z$ range, the numbers of red and blue galaxies were first counted. Each of the red, blue, and (red+blue) sam-ples was pared to $N_{small}$ galaxies, where $N_{small}$ is the smaller of red or blue, by choosing galaxies at random from among them. Then the separations, $r$, for all combinations of RR, BB, and RB galaxies were binned in sixty 0.00025 wide bins between $r=0$ and $r=0.015$. Occasionally in the SDSS data nearby galaxies appear more than once with different IDs when different points in the same galaxy are chosen as centers. Therefore pairs with separations $<1\times10^{-5}$ were excluded in order to remove possible duplicates. (This is several times the radius of a typical large galaxy.) The correlation functions were calculated as in Eq. (3) using the pair counts in each separation bin.

## 4. RESULTS

Typical examples of the correlations *vs.* separation are shown in Figure 2. The filled red circles are the $\xi_{RR}$, the blue open circles in Fig. 2a are the $\xi_{BB}$, and the asterisks are the $\xi_{RB}$. Error bars are usually smaller than the points. As defined in Eq. (3), $\xi_{RR}$ and $\xi_{BB}$ are mirror images of each other, so the $\xi_{BB}$ are only shown for the top plot. The red galaxies show a strong positive correlation with each other for separations <0.003 with an overwhelming statistical significance. The red and blue galaxies are very strongly anticorrelated with each other. For larger separations the $\xi$ are governed by the relative spatial densities of the red and blue galaxies in the chosen volume. For separations > 0.008 the correlations are roughly consistent with 0 over the whole redshift range studied. This shows that the technique of choosing equal numbers of red and blue galaxies in the volume works very well in removing systematic effects due to the spatial density of observed galaxies decreasing with distance. The amplitudes of the correlation functions at small separations were found to be remarkably consistent over the entire redshift range, further evidence for the power of the red/blue comparison technique.

The first separation bin is typically anomalous and was not used in the calculation of the correlation lengths. This is primarily due to the difficulty in resolving distinct galaxies that are close to each other on the sky and the masking of distant galaxies due to overlap with nearer ones. It has less effect on the red-blue correlation since it is easier to distinguish galaxies that are close to each other if they have different spectra. At larger $z$ there is some bias due to gravitational lensing. For larger $z$ there is also some smearing of the $\xi$ at small separations due to the uncertainties in the SDSS redshift determinations. This was minimized by removing the few galaxies for which the uncertainty in $z$ is $>0.02\,z$; it is only important for the larger $z$ slices. The contribution to the errors in the correlation lengths due to smearing is discussed below.

There is no compelling model for the variation of $\xi$ with $r$. The definition of the correlation length $L_0$ is therefore somewhat arbitrary. As an *ad hoc* model-independent definition we use the mean $r$ over redshift bins 2 through 15 weighted by $\xi$ for that separation bin, $L_0 \equiv \sum_{2}^{15} \xi r / \sum_{2}^{15} \xi$. This corresponds to separation bin centers in the range 0.000375 to 0.003875. The statistical uncertainties in the $L_0$ were calculated from the statistical errors in the 14 bins as given by Eq. (4). The statistical errors were usually much smaller than the total error (Table I). The length $L_0$ is a measure of the average visual cluster size, which is determined by the dark matter potential well the galaxies reside in, and is a measure of the average radius of clusters in the redshift slice.

TABLE I. $U$-$Z$ is the cut used to distinguish red and blue galaxies, $\bar{\xi}$ is the average correlation amplitude in bins 2–15, $L_0$ Corr is the correlation length after all corrections, StatErr is its statistical error, and TotErr its total error.

| Redshifts | $U$-$Z$ | $\bar{\xi}$ | $10^3$ $L_0$ Corr | StatErr | TotErr |
|---|---|---|---|---|---|
| 0.01-0.02 | 1.923 | 0.168 | 1.263 | 0.016 | 0.130 |
| 0.02-0.04 | 2.395 | 0.220 | 1.507 | 0.004 | 0.021 |
| 0.04-0.06 | 2.773 | 0.188 | 1.571 | 0.006 | 0.027 |
| 0.06-0.08 | 3.094 | 0.196 | 1.610 | 0.005 | 0.013 |
| 0.08-0.10 | 3.299 | 0.199 | 1.609 | 0.007 | 0.029 |
| 0.10-0.15 | 3.510 | 0.166 | 1.592 | 0.008 | 0.016 |
| 0.15-0.20 | 3.818 | 0.174 | 1.630 | 0.019 | 0.031 |
| 0.20-0.30 | 3.500 | 0.245 | 1.677 | 0.045 | 0.060 |
| 0.30-0.50 | 4.000 | 0.233 | 1.728 | 0.077 | 0.084 |
| 0.50-0.75 | 3.500 | 0.392 | 2.100 | 0.400 | 0.504 |

## 5. UNCERTAINTIES AND SYSTEMATICS

The correlation lengths were found to be very robust relative to changes in the selection criteria. Many checks of the code were made. In particular, when the redshifts were scrambled at random among the galaxies the correlations disappeared entirely.

Due to the large number of galaxies in the sample, it was possible to study possible systematic effects and estimate the resulting uncertainties using the data itself. Generally the uncertainty estimates themselves were limited by the statistical error in the determination of $L_0$. These uncertainties are discussed in the following. They were combined in quadrature with the statistical error to get the total error given in Table I.

*Red-Blue Separation.*–The $U$-$Z$ values for the red-blue separation are given in Table I. The uncertainties were estimated by varying the cut by at least 10% from the value there. They were comparable to the statistical errors with no apparent systematic effect with increasing $z$.

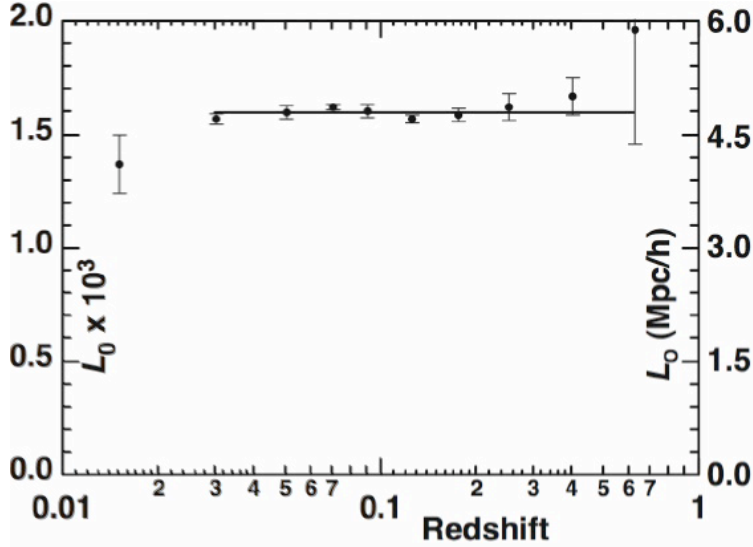

FIG. 3  $L_0$ vs. redshift after all corrections.  The line is a fit to a constant, excluding the first point, that gives  $\bar{L}_0 = (1.599 \pm 0.008) \times 10^{-3}$ with a $\chi^2$/dof of 1.22.  The scale on the right gives $L_0$ in units of Mpc/h.

*Smearing due to zerr.–* The smearing in separations due to the uncertainty in the measured $z$ tends to decrease $\xi$ at small $r$.  This effect was studied by smearing the separations by an additional ±50% of *zerr*.  This procedure could not be used for the 3 highest $z$ bins because the statistical errors in the $\xi$ were too large; instead the additional smearing was varied until $\xi$ in the second bin decreased to half its maximum value.  This error was always less than the statistical.

*Edge effects.–* Galaxies at the edge of the defined volumes, both the red and blue, will have neighbors on one side only.  Possible edge effects were eliminated by looking at correlations between galaxies in a "box" and those in a larger box that was at least 0.005 larger on all sides.  As a check, when the box sizes were made the same, there was no change in the $L_0$ within statistical errors.  This was not possible with the 0.01<z<0.02 slice which is effectively all "edge".  The uncertainty in $L_0$ for this slice was estimated as ±0.12 by varying the boundaries of the two boxes; this was the dominant error for that slice.  Despite the piecemeal coverage in $\delta$ of the hemisphere toward $RA=0°$, the $L_0$ were found to be the same as the $RA\sim180°$ hemisphere within statistical errors that were about 5 times larger.

*Variation of $L_0$ with absolute magnitude.–* A galaxy at larger $z$ has to have a higher (less negative) absolute magnitude $M$ to be observed. Thus there are fewer galaxies per unit volume in the sample at larger $z$. The pair separations therefore tend to increase with $M$ and $z$ for both the red and blue samples. The effect on $L_0$ is small because we are effectively comparing the separation of red galaxies to blue and we chose equal numbers of red and blue galaxies (see Sec. 2 and 3), so that their average spatial densities were the same. It was not possible to evaluate any residual bias in the individual redshift slices because the range of absolute magnitudes was too small. Therefore this correction had to be determined using the complete sample.

The average $M$ in the green band for each $z$ slice was found to vary linearly with $\log z$ from $-16.6$ for $0.01<z<0.02$ to $-20.63$ at $0.20<z<0.30$. The galaxies in each $z$ slice were divided into two halves with $M>\bar{M}$ and $M<\bar{M}$, where $\bar{M}$ is the average $M$. Then $L_0$ was calculated for each half and the slope $\Delta L_0/\Delta M$ determined. The absolute magnitudes were determined using the apparent magnitude and $K$ corrections.[3] An error-weighted fit to a constant $\Delta L_0/\Delta M$ over the ten $z$ slices gave an average $\langle \Delta L_0/\Delta M \rangle = 0.0405\pm0.0114$ with the error adjusted to give a $\chi^2/dof = 1$. Adjustments to the $L_0$ were then made using the value for the $0.08<z<0.10$ slice as an anchor, $\langle \Delta L_0/\Delta M \rangle$ given above, and the difference in $\bar{M}$ for that $z$ slice and the $0.08<z<0.10$ slice. As a further check on the effect of decreasing spatial density with $z$, runs were made with only 25% of the galaxies that were chosen at random. At least 16 runs were made for each $z$ slice to reduce the statistical errors in $L_0$. The averaged $L_0$'s for these runs was consistently 3.6% higher (within statistical errors) than the normal runs for all $z$.

*Variation of $L_0$ with cosmology.–* With $q_0 = 0.5$, $L_0$ changed about -2.5% at $z=0.2$.

6. CONCLUSIONS

The corrected $L_0$ are given in Table I along with error estimates and are plotted in Fig. 3. The horizontal line shows a least squares fit to a constant $L_0$ that yields $\bar{L}_0 = (1.599 \pm 0.008) \times 10^{-3}$ with a $\chi^2/\text{dof} = 1.22$. The first point is problematical because of the small volume involved, and it was not used in the fit. Using $H_0 = 100h$ km s$^{-1}$Mpc$^{-1}$ gives $\bar{L}_0 = 4.797 \pm 0.024$ Mpc/h. This result is consistent with other data on cluster size[3], though the sizes are definition dependent. There is no evidence for evolution of cluster size with redshift. Conversely there is no sign of any variation of the Hubble expansion rate with $z$. A more careful study of the correlations at small $z$ and comparison with other standard rulers is needed to make an absolute calibration of the low end of the distance scale.

The technique of using the red *vs.* blue galaxy comparison to determine average cluster sizes has the advantages of great statistical power as well as being almost free of selection biases. I find that the average cluster size can serve as a cosmic ruler with an accuracy ~2% out to redshifts of 0.5. With deeper surveys and more careful redshift measurements it will be possible to extend it to considerably larger $z$. It will also be possible to make much finer grained studies of average cluster sizes out to $z \sim 0.5$, which will enable tests of the cosmological principle and of the standard cosmological model.

The amplitude of the correlation ($\bar{\xi}$ in Table I) is a measure of the clustering strength and is remarkably constant over the whole $z$ range. This shows that the cluster number density remains substantially constant out to $z \sim 0.5$ and may actually increase.